\documentclass[iop]{emulateapj}
\slugcomment{ApJ Letters, accepted}

\shorttitle{The Formation of Thick Disks in Galaxies}
\shortauthors{Lehnert et al.}

\begin{document}

\title{The Milky Way as a High Redshift Galaxy: The Importance of
Thick Disk Formation in Galaxies}

\author{Matthew D. Lehnert\altaffilmark{1}, Paola Di Matteo\altaffilmark{2}, Misha Haywood\altaffilmark{2}, \& Owain N. Snaith\altaffilmark{3}}

\affil{\altaffilmark{1}Institut d'Astrophysique de Paris, UMR 7095,
CNRS, l'Universit{\'e} Pierre et Marie Curie, 98 bis boulevard Arago,
75014 Paris, France; email: lehnert@iap.fr}

\affil{\altaffilmark{2}GEPI, Observatoire de Paris, UMR 8111, CNRS, Universit\'e
Paris Diderot, 5 place Jules Janssen, 92190 Meudon, France}

\affil{\altaffilmark{3}Department of Physics \& Astronomy, University
of Alabama, Tuscaloosa, Alabama, USA}

\begin{abstract}
We compare the star-formation history and dynamics of the Milky Way (MW)
with the properties of distant disk galaxies. During the first $\sim$4
Gyr of its evolution, the MW formed stars with a high star-formation
intensity (SFI), $\Sigma_{\rm SFR}\approx$0.6 M$_{\sun}$ yr$^{-1}$
kpc$^{-2}$ and as a result, generated outflows and high turbulence
in its interstellar medium.  This intense phase of star formation
corresponds to the formation of the thick disk. The formation of the
thick disk is a crucial phase which enables the MW to have formed
approximately half of its total stellar mass by z$\sim$1 which is
similar to ``MW progenitor galaxies'' selected by abundance matching.
This agreement suggests that the formation of the thick disk may be
a generic evolutionary phase in disk galaxies.  Using a simple energy
injection-kinetic energy relationship between the 1-D velocity dispersion
and SFI, we can reproduce the average perpendicular dispersion in stellar
velocities of the MW with age.  This relationship, its inferred evolution,
and required efficiency are consistent with observations of galaxies
from z$\approx$0$-$3.  The high turbulence generated by intense star
formation naturally resulted in a thick disk, a chemically
well-mixed ISM, and is the mechanism that links the evolution of MW to
the observed characteristics of distant disk galaxies.
\end{abstract}


\section{Introduction}\label{sec:intro}

The physical processes that regulate the growth and evolution of galaxies
are both complex and myriad \citep[e.g.][]{hopkins13}.  There will
likely be no single methodology that enables us to unravel the
physics underpinning the variety of phenomena we observe in galaxies
as a function of epoch and galactic environment.  However, one avenue
which may allow us to obtain a deeper understanding of these processes
is to use the findings of {\it in situ} galaxy studies -- observing how
the ensemble of galaxies changes with look-back time -- and galactic
archeology -- the unraveling of the detailed star formation, dynamical,
and chemical and elemental history through observation of individual stars
or simple stellar populations (e.g. globular clusters) in nearby galaxies.
Such analyses, as our knowledge from the direct study of galaxies over a
wide range of redshifts and galactic archeology grows, become more robust.

That is the purpose and novelty of this paper, we use the star
formation history (SFH) and dynamical evolution of stars in the Milky Way
\citep{haywood13, snaith14} to connect quanitatively its cosmic evolution
to the observed dynamical evolution and the physical characteristics of
high-redshift disk galaxies.  Specifically, we will compare the stellar
mass growth and star-formation rates (SFR), star-formation intensity
(SFI), and the stellar and gas kinematics of MW, to the dynamical and
star formation properties of distant disk galaxies. In \citet{haywood13},
we proposed that the formation of the thick disk represented an important
phase of the evolution of the MW both in its stellar mass growth and its
influence over subsequent evolution of the MW.  But can we physically
link this phase to what is known about distant galaxies?

Star-forming galaxies at high redshift accumulate their stellar masses in
a systematic way as seen in the evolving ridge lines in the star-formation
rate-stellar mass plane \citep[SFR-M$_{\star}$, e.g.][]{daddi07,
elbaz07, reddy12}.  \citet{snaith14} showed that the MW has a similar
SFH as that of possible MW-like progenitor galaxies selected by abundance
matching \citep{vandokkum13} or by their evolution in the SFR-M$_{\star}$
plane \citep{patel13}.  Does the MW accumulate its stellar mass in the
same way as distant galaxies do in the SFR-M$_{\star}$ plane?

Comparisons between the evolution of MW progenitors and the
dynamical evolution and growth of the MW, as inferred from the dynamics
and elemental abundances of stars in the solar vicinity, allows us
to link the properties of the MW to the physical conditions of
the interstellar medium (ISM) in distant galaxies. Disk galaxies at
high redshift are known to be highly turbulent \citep[e.g.][]{L09}.
Broad spatially-resolved line emission from the warm ionized and cold
molecular gas are observed in distant galaxies \citep[e.g.][]{L09,
swinbank11, MD13}. \citet{L09, L13} suggested that the broad lines
observed are a result of turbulence and bulk motions generated by the
mechanical and radiative energy output from the intense star formation
occurring within these galaxies. If the MW formed a significant fraction
of its stellar mass at 1$<$z$<4$ via similarly intense star formation,
then there must be a fossil record of this vigorous energy injection in
the kinematics and elemental abundances of its individual stars.
By quantifying the relationships between the stellar and gas dynamics
and SFI at different epochs for the MW, we suggest that this fossil
record is in the thick disk, and that indeed thick disks must be a generic
evolutionary phase of disk galaxy formation, representing the imprint
of the intense star formation observed in disk galaxies at high redshift.

\section{The Milky Way as an evolving star-forming galaxy}\label{sec:SFRmass}

\begin{figure}
\includegraphics[width=8.8cm]{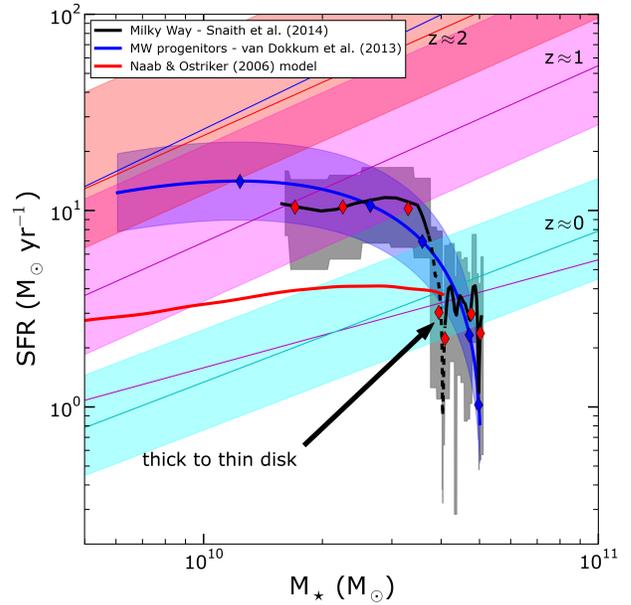}
\caption{The relationship between the stellar mass, M$_{\star}$
(M$_{\sun}$), and SFR (M$_{\sun}$ yr$^{-1}$).  The evolution of the SFH
of the MW \citep[black line, dashed region represents the transition of
thick to thin disk; ][]{snaith14}; specific points in the MW's evolution
with redshift are indicated (red diamonds; z=4, 3, 2, 1.3, 0.8, 0.3,
0.1 from left to right).  The grey region represents the uncertainties
in the SFR of the MW. For comparison, we show the SFR-M$_{\star}$
ridge lines and scatter (shaded regions) from \citet[][z$\approx$2;
red]{daddi07}, \citet[][z$\approx$2; blue]{reddy12}, \citet[][z$\approx$0
and 1; cyan and magenta respectively]{elbaz07}, and \citet[][z$\la$0.2;
magenta]{lara-lopez13}.  We also show the evolution of MW progenitor
galaxies (blue line) over the redshift range 2.5--0 decreasing from
left to right, with individual redshifts, z=2, 1.3, 0.8, 0.3, 0.1,
indicated by blue diamonds \citep[][blue shaded region indicates
a $\pm$0.2 dex diversity in the SFR; see also \citeauthor{patel13}
\citeyear{patel13}]{vandokkum13}. Within the uncertainties, the evolution
of the MW and possible progenitors are similar (see text for details).
Also shown is the model of \citet[][red line]{naab06}.}
\label{fig:SFRevol}
\end{figure}

To investigate how the MW evolved compared to galaxies observed at
various epochs, we use the SFH and velocity dispersions of stars
from \citet{haywood13} and \citet{snaith14}.  \citet{snaith14} used a
closed-box model to iteratively fit the distribution of individual stars
in the galaxy in the [$\alpha$/Fe]-age plane (giving most of the weight
to [Si/Fe]).  They found that the growth of the MW follows two regimes
of nearly constant star-formation rates. The first regime corresponds to
the formation of the thick disk. During this epoch the Galaxy sustained
a SFR$\sim$10-15~M$_{\odot}$ yr$^{-1}$ and formed about half of its
current stellar mass. The second regime occured during the growth of
the thin disk which is characterized by lower star-formation rates,
$\sim$ few M$_{\odot}$ yr$^{-1}$, and formed the other half of the MW's
current stellar mass.  The change from the thick to thin disk formation
occurred at z$\approx$1--1.4 when there is an abrupt drop in the star
formation rate (Fig.~\ref{fig:SFRevol}). 

\subsection{Evolution in the SFR-M$_{\star}$ plane}

Using the SFH of MW, we can determine how the evolution of the MW
compares with the ensemble of galaxies in the SFR-M$_{\star}$ plane.
Star-forming galaxies form ridge lines in the SFR-M$_{\star}$ plane
\citep[e.g.][]{elbaz07, daddi07}.  These ridge lines
imply that galaxies accumulate their stellar mass in a systematic way as a
function of stellar mass.  The growth history of the MW is similar to that
of abundance-matched MW progenitors \citep{vandokkum13}, with about half
of their total stellar mass formed by z$\sim$1 during which time the SFR
is a factor of 3-4 higher than that at z$\la$1 (Fig.~\ref{fig:SFRevol}).
The contribution to the mass growth by the bulge is likely to be small
at z$\ga$1 \citep{shen10, dimatteo14} and it is the growth of the thick
disk that accounts for similarities in the stellar mass growth of the
MW and its antecedents.  While at z$\la$1, the SFR and stellar masses
of the MW and its progenitors are strikingly similar, at z$\ga$1, the
MW appears to grow significantly earlier than its possible progenitors.
However, the uncertainties in the estimated ages of stars in the MW,
especially the oldest stars, upon which its SFH is based \citep{snaith14}
and both the diversity of and the uncertainty in the SFH of abundance
matched galaxies, means that the two SFH are comparable \citep[Fig.~2
in][]{snaith14}. Specifically, the uncertainties in the ages of the oldest
stars of the MW are $\sim$1 Gyr (10\% uncertainty) result
in large differences in the corresponding redshifts (the
difference in lookback time between redshifts 2 and 4 is $\sim$1 Gyr).

\begin{figure}
\includegraphics[width=8.8cm]{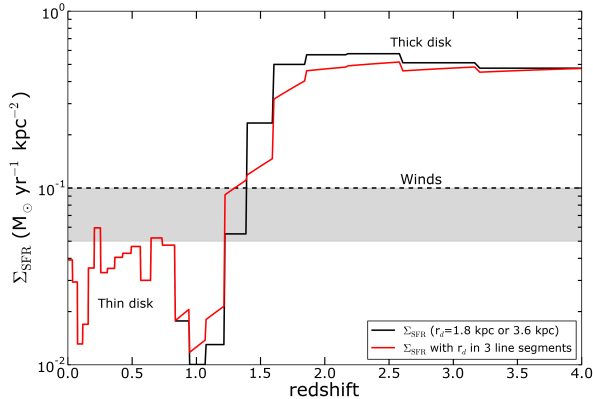}
\caption{The evolution of the MW's star-formation intensity, $\Sigma_{\rm
SFR}$ (M$_{\sun}$ yr$^{-1}$ kpc$^{-2}$), as a function of redshift.
$\Sigma_{\rm SFR}$ was estimated using the MW's SFH and assuming a disk
scale length, r$_{\rm d}$=1.8 kpc for look-back times between 9-14 Gyr
and r$_{\rm d}$=3.6 kpc at later times.  These two regimes are labeled
``thick disk'' and ``thin disk''. To demonstrate the impact of a smooth
transition in the growth of the disk from thick to thin, we made a three
segment time evolution of the disk scale length by linearly connecting
3 epochs of disk evolution, the thick disk phase, r$_{\rm d}$ increases
from 1.6 to 2.0 kpc from ages of 13.7 to 10 Gyrs, a transition phase
from 2.0 to 3.6 kpc over ages of 10 to 7 Gyrs, and then constant size
evolution, r$_{\rm d}$=3.6 kpc, for ages younger than 7 Gyrs (red line;
we use this evolution subsequently in our analysis).  We also indicate
the region of SFI where starburst-driven outflows
become prevalent in nearby galaxies \citep{heckman03}.}
\label{fig:SFIvsz} 
\end{figure}

Even allowing for such shifts in redshift, two things are clear:
The position of the MW in the SFR-M$_{\star}$ plane lies below
the ridge lines over z=0-3 and the stellar mass growth rate at
early times is more rapid than predicted by gas accretion models
\citep[e.g.][Fig.~\ref{fig:SFRevol}]{naab06,pipino13}.  These models
significantly underestimate the impact of the formation of the thick
disk and the total stellar mass formed during this epoch.

\subsection{Star-formation intensity evolution}\label{subsec:SFIevol}

During its formation, the thick disk had a short radial scale-length
\citep{bovy12} and a high SFR and thus likely a high SFI.  If the
star formation is sufficiently intense, the young MW may have
generated outflows \citep{lehnert96}.  Outflows from galaxies play a
crucial role in establishing relationships among galaxies such as the
mass-metallicity relationship, the metallicity-radius relation, and
many others \citep{heckman00}.  To understand the role of the energy
output of massive stars in determining the properties of the MW, we
estimate the evolution of its star-formation intensity ($\Sigma_{\rm
SFR}$=SFR/2$\pi$r$_{\rm d}^2$; Fig.~\ref{fig:SFIvsz}).  Currently, there
is no known way to connect the change in the radial scale length of the
MW's disk to stellar ages and thus its size evolution.  The simplest model
consistent with the data is one where the thick disk has a scale length,
r$_{\rm d}$=1.8 kpc, and the thin disk has a radial scale length, r$_{\rm
d}$=3.6 kpc \citep{bensby11, bovy12}. The transition between the thick
and thin disk occurs at the end of the thick disk phase \citep[$\sim$9
Gyr ago, Fig.~\ref{fig:SFIvsz};][]{haywood13}.  Before discussing the
implications of the MW's resulting SFI, we emphasize that changing
the functional form for the temporal evolution of the MW's scale
length does not change SFI(z) significantly.  Selecting an evolution
with a smooth transition from thick to thin disk, for example,
a 3-segment evolution of the scale length, shows that adopting
different time-dependent r$_{\rm d}(z)$ do not significantly change the
overall evolution of $\Sigma_{\rm SFR}$ (Fig.~\ref{fig:SFIvsz}) -- the
break in the SFI intensity was predominately due to the dip in the SFR.

Irrespective of the precise form of the radial-scale evolution of the MW
disk, the MW was likely driving outflows and generating significant levels
of turbulence during the formation of its thick disk.  The SFI of the MW
during it thick disk formation phase lies above that required to generate
strong outflows in the local universe \citep[Fig.~\ref{fig:SFIvsz};
e.g.][]{lehnert96}.  This finding is fundamental to understanding the
early evolution of the MW: the structure and kinematics of the thick
disk, characterized by large scale heights and high velocity dispersion
of its stellar population \citep{bovy12, haywood13}, is likely a direct
result of the intense star formation generating a highly turbulent ISM.
Importantly for the chemical evolution of the MW and the lack of radial
metallicity gradients found in the MW \citep{cheng12}, high levels
of turbulence will naturally lead to a vigorous mixing and weakening,
perhaps destroying, any intrinsic gradients in the disk.  A vertical
metallicity gradient may develop if the thick disk had progressively
lower scale height and became more metal-rich as it grew
\citep{haywood13}.  A decreasing scale height is expected due to the
increasing stellar surface and mass density of the disk.  Subsequently,
the MW evolved rather quiescently, its star-formation intensity was low,
which would naturally lead to lower dispersions in the gas and stars
forming a thin disk \citep[e.g.][]{dib06,L13}.  These results suggest
a direct link between the conditions of the ISM in the evolving MW and
the formation of its thick and thin disks.

Interestingly, the MW disk beyond 10~kpc has properties different from
those of the inner thin disk, in particular a break in metallicity,
\citep[a situation observed in other galaxies;][]{bresolin12},
but has similar [$\alpha$/Fe] as the young thick disk \citep{haywood13}.
Although the inner and outer disks seem otherwise rather discontinuous,
outflows are a plausible explanation for the similar abundance patterns
suggesting that not all the outflowing material escaped.

\subsection{Velocity dispersion evolution of the stars and gas: Relation
to the SFI}

\citet{L09} and \citet{L13} proposed that the spatially-resolved broad
(50-250 km s$^{-1}$) optical emission line gas observed in distant
(z$\approx$2) galaxies are due to high levels of turbulence and bulk
flows generated by the energy output of young stars.  Furthermore, they
argued that the high turbulence observed in the warm ionized medium (WIM)
is likely captured in the cold molecular gas by a mass and energy
flow. They characterized this phenomenon as a relation between the gas
velocity dispersion and the SFI, $\sigma$=$\epsilon$$\sqrt{\Sigma_{\rm
SFR}}$. This relationship is a simple formulation of the coupling
(with efficiency, $\epsilon$) between the energy injection into
the ISM and the kinetic energy of gas: specifically, $\sigma_{\rm
1-D}^2$=$\epsilon^2$$\Sigma_{\rm SFR}$/3 + $\sigma_{\rm 0, 1-D}^2$,
where $\sigma_{\rm 0, 1-D}$ accounts for dispersions which are not
associated directly to star formation \citep[][]{agertz09,stilp13}.
Stars would then form in a population of molecular clouds with a high
relative velocity dispersion embedded in a thick gas disk. Since the
stellar population formed out of such gas, it would have a similarly
high velocity dispersion and be geometrically thick.

\begin{figure}
\includegraphics[width=8.8cm]{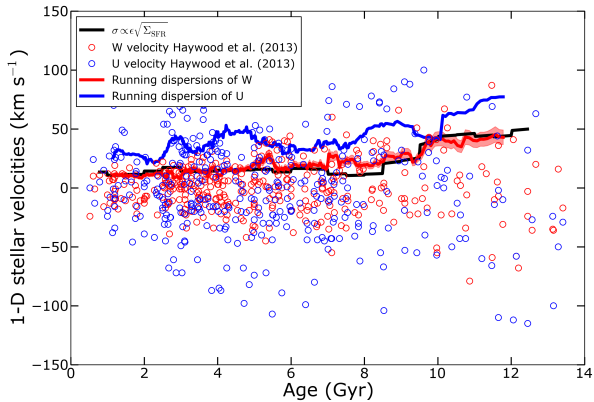}
\caption{The 1-D velocities of stars in the MW as a function of their
age \citep[red circles indicate the W-component -- velocities out of the
plane -- and blue circles indicate the U-component -- radial velocity
component, see][]{haywood13}.  We also indicate the relationship of
the form, $\sigma_{1D}^2$=($\epsilon$ $\sqrt{\Sigma_{\rm SFR}})^2$/3
+ $\sigma_{\rm 0, 1-D}^2$ (black line, see text for details) and the
running dispersion in the W- and U-velocities of individual stars in
the MW (red and blue lines respectively and the dispersion of the running
dispersion of the W-component as the red shaded region).}
\label{fig:dispvsage} 
\end{figure}

Having estimated the SFI evolution of the MW (\S~\ref{subsec:SFIevol}),
we can use the relation between $\sigma$ and $\Sigma_{\rm SFR}$ to
estimate the dispersion in stellar velocities in the solar neighborhood
\citep{haywood13}.  Such a relation reproduces the trends in the evolution
of the dispersions in the W-velocity (perpendicular to the plane) and
U-velocity (radial) components of individual stars in the MW if we adopt
$\epsilon$=110 km s$^{-1}$ (M$_{\sun}$ yr$^{-1}$ kpc$^{-2}$)$^{-1/2}$ and
$\sigma_{\rm 0, 1-D}$=5-10 km s$^{-1}$ (Fig.~\ref{fig:dispvsage}). The
coupling efficiency required to explain the MW data is
similar to that necessary for the SFI-$\sigma$ relation in high-redshift
galaxies \citep{L13}.  Radial velocities may be affected by secular
processes, which will increase the radial motions over time.  Thus,
the dispersion in the velocities in the radial direction may be taken
as an upper limit to the initial dispersion.  The vertical velocities
may be affected by heating through satellite accretion, for example,
but our analysis indicates that the $\sigma$-age relation is close
to the initial one and that subsequent dynamical processes perhaps
modified it, but did not erase it.  The evolution in the dispersion of
velocities among stars in the MW is reminiscent of the general decrease
in spatially-resolved line widths with decreasing redshift observed in
high-redshift galaxies \citep{epinat10, kassin12}.

\begin{figure}
\includegraphics[width=8.8cm]{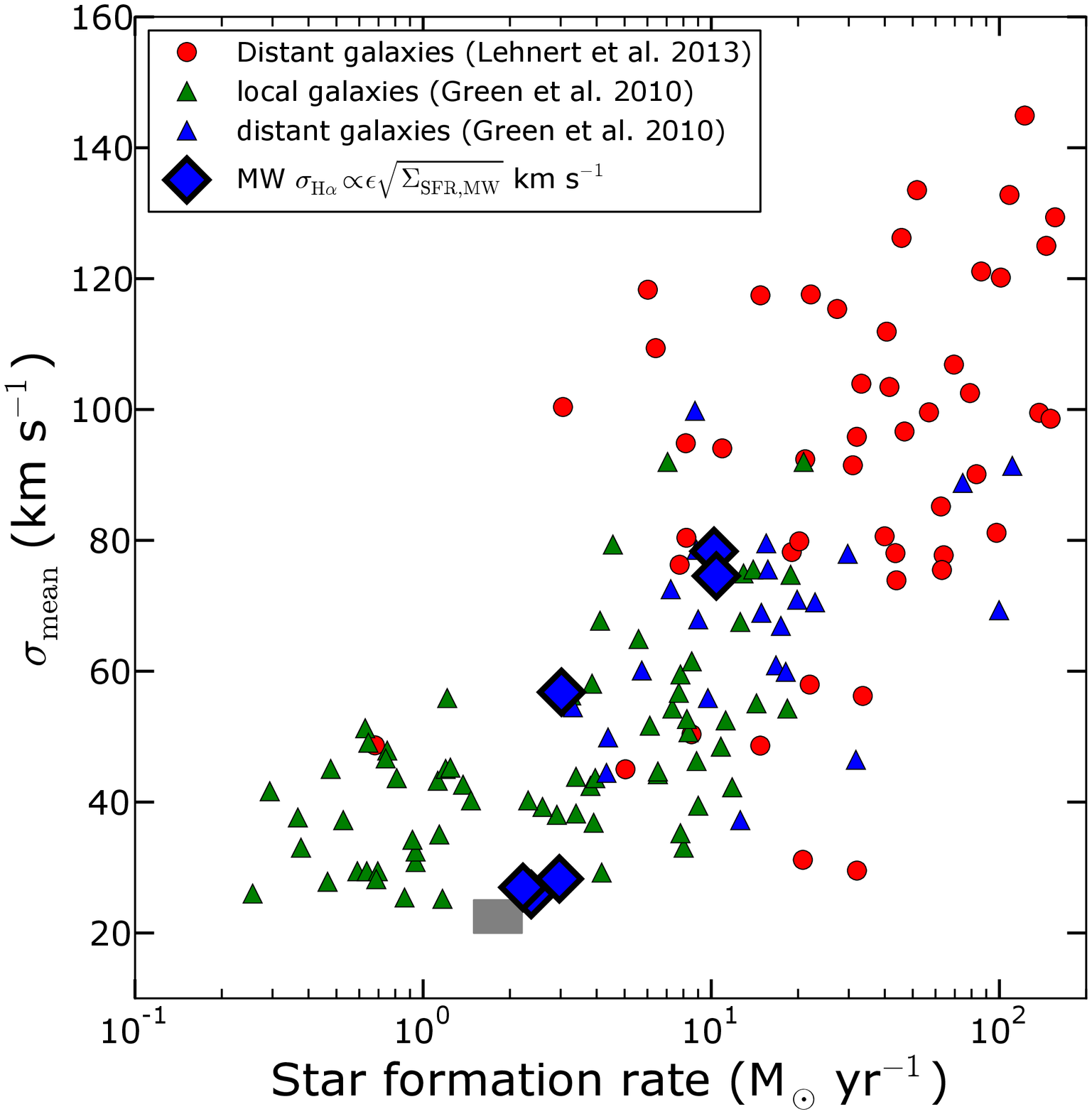}
\caption{The mean velocity dispersion of the H$\alpha$ line profile,
$\sigma_{\rm mean}$ (km s$^{-1}$), versus the integrated 
SFR (M$_{\sun}$ yr$^{-1}$) for nearby and distant galaxies \citep{L13,
green10}.  We also indicate, using the SFH of the MW \citep{snaith14},
where the MW would lie in the relationship assuming $\sigma_{\rm
H\alpha, mean}^2$=($\epsilon$ $\sqrt{\Sigma_{\rm SFR}})^2$ + $\sigma_{\rm
0,thermal}^2$ (blue diamonds which represent from left to right, z=0.1,
0.3, 0.8, 1.3, 2, 3).  We used a standard relation to convert between
total H$\alpha$ luminosity and SFR for both the local and distant
galaxies \citep[appropriately scaled for the IMF; ][]{kennicutt98}.
The grey box shows the range of current SFR estimated for the MW
\citep[1.9$\pm$0.4 M$_{\sun}$ yr$^{-1}$; ][]{chomiuk11} and velocity
dispersions of the H$\alpha$ line along over 100 lines-of-sight in the
Milky Way \citep[$\approx$20-25 km s$^{-1}$; ][]{haffner09}.}
\label{fig:dispvsSFR} 
\end{figure}

We have connected the evolution of the velocity dispersion of the disk
stars of the MW to a relationship observed for gas in distant galaxies,
but what about the general population of star-forming galaxies over a
wide range of redshifts?  Since we cannot typically measure the stellar
velocity dispersions of distant disk galaxies, we will use the H$\alpha$
line widths for this comparison. Similar to the relation between
the SFI and the gas velocity dispersion, there is also a relationship
between the total SFR and the mean gas velocity dispersion, $\sigma_{\rm
mean}$, in local and distant star-forming galaxies \citep{green10, L13}.
This relationship is determined using the velocity dispersion and SFR
estimated from the width and luminosity of H$\alpha$ line and it is
appropriate for the WIM, not the stars (cf.  Fig.~\ref{fig:dispvsage}).
We demonstrate that the MW was similar to local and distant disk galaxies
(Fig~\ref{fig:dispvsSFR}) if we assume the relationship, $\sigma_{\rm
H\alpha, mean}^2$=($\epsilon$ $\sqrt{\Sigma_{\rm SFR}})^2$ + $\sigma_{\rm
0,thermal}^2$, with $\epsilon$ = 110 km s$^{-1}$ (M$_{\sun}$ yr$^{-1}$
kpc$^{-2}$)$^{-1/2}$ and $\sigma_{\rm 0, thermal}$=18.5 km s$^{-1}$, where
$\sigma_{\rm 0, thermal}$ is the thermal velocity of Hydrogen atoms at
7500 K, the approximate temperature of the diffuse WIM in the Milky Way
\citep{haffner09}.  We are using the 3-dimensional velocity dispersion
in estimating $\sigma_{\rm mean}$ for the MW because these observations
sample galaxies with a variety of inclinations and gas disk thicknesses.
The SFR and dispersions are estimated for intensely star-forming galaxies
from z$\approx$1-3 \citep[][and references therein]{green10, L13},
not necessarily progenitors of the MW; and for local, z$\la$0.1, disk
galaxies \citep[see][]{green10}.  Our estimates are consistent with both
the MW's current SFR \citep{chomiuk11} and H$\alpha$ velocity dispersions
\citep{haffner09} and those estimated for galaxies with z=0--3: the MW's
ISM evolved like that of the general population of galaxies which was shaped
by the SFI and SFR.

\section{Discussion and Conclusions}\label{discussion}

Our results imply that the thickness of the disk in the early evolution
of the MW was related to its SFI, total SFR, and the conditions of
the ISM.  This relationship may be direct in that the mechanical and
radiative energy output from the young massive stars may drive a simple
energy injection--kinetic energy relationship or it could be indirect
in that other processes may contribute, such as gas accretion \citep[but
see][]{hopkins13}.  Such a relation provides a natural explanation as to
why the disk is both thick and the stars within it have a high dispersion
in velocity and why the oldest stars in the MW show no metallicity
gradient. The high turbulence levels weaken any intrinsic metallicity
gradients due to the short mixing timescales of the gas.

Our analysis implies a central role for the formation of the thick
disk in the evolution of disk galaxies generally: the only way the MW
progenitors match the evolution of the MW is through the formation of
thick disks during their early evolution. It is the relationship between
star formation and turbulence that draws a direct connection between
observations of distant galaxies and the inferred evolution of the MW.
The formation of the thick disk is simply a reflection of the fact that
galaxies are gas rich and have intense star formation in their early
evolution and it is the impact of high SFI, which results in the disk
being thick, that links the early evolution of the MW to the MW-like
galaxy progenitors.  Models which result in a constant SFR with no
significant increase at early times \citep[e.g.][]{naab06}, are at odds
with what has been inferred for both MW progenitors and the MW itself.
The fossil record of the MW and observations of MW progenitor galaxies
imply that the formation of the thick disk was a generic process in the
growth of disk galaxies and that the thick disk represents the imprint
of intense star formation.

\begin{acknowledgements} 
The authors acknowledge support from the ANR under contract
ANR-10-BLAN-0508.  Support for ONS was provided by NASA
Research grant HST-AR-12837.01-A.
\end{acknowledgements}

\end{document}